\documentclass[11pt,twoside]{article}
\usepackage{asp2010}

\resetcounters
\aspvolume{455}
\aspcpryear{2012}
\aspvoltitle{4$^{th}$ {\em Hinode} Science Meeting: Unsolved 
Problems and Recent Insights}
\aspvolauthor{L.~R.~Bellot Rubio, F.~Reale, and M.~Carlsson, eds.}

\begin{document}

\title{Nanoflare Evidence from Analysis of the X-Ray Variability of 
an Active Region Observed with {\em Hinode}/XRT}
\author{S.~Terzo,$^1$ F.~Reale,$^{1,2}$ M.~Miceli,$^{1,2}$ R.~Kano,$^3$  
S.~Tsuneta,$^3$ and J.~A.~Klimchuk$^4$}
\affil{$^1$Dipartimento di Scienze Fisiche ed Astronomiche, Universit\`a
	             degli studi di Palermo,
              Via Archirafi 36, 90123, Palermo, Italy}
\affil{$^2$INAF Osservatorio Astronomico di Palermo,
	     Piazza del Parlamento 1, 90134 Palermo, Italy}
\affil{$^3$National Astronomical Observatory of Japan,
	     2-21-1 Osawa, Mitaka, Tokyo, 181-8588, Japan}
\affil{$^4$NASA Goddard Space Flight Center, 
	     Greenbelt, MD 20771, USA}

\begin{abstract}
The heating of the solar corona is one of the big questions in astrophysics. Rapid pulses called nanoflares are among the best candidate mechanisms. 
The analysis of the time variability of coronal X-ray emission is potentially a very useful tool to detect impulsive events. We analyze the small-scale variability of a solar active region in a high cadence {\em Hinode}/XRT observation. The dataset allows us to detect very small deviations of emission fluctuations from the distribution expected for a constant rate. We discuss the deviations in the light of the pulsed-heating scenario.
\end{abstract}

\section{Introduction}
The investigation of the heating mechanisms of the confined coronal
plasma is still under intense debate. It is widely believed that the
energy source for coronal heating is the magnetic energy stored in the
solar corona. An unsolved problem is how this magnetic energy is
converted into thermal energy of the confined coronal plasma. As
\citeauthor{1988ApJ...330..474P} proposed in 1988, rapid pulses called
\emph{nanoflares} are among the best candidate mechanisms of magnetic
energy release.  Nowadays a challenging problem is to obtain evidence
that such nanoflares are really at work.  If small energy discharges
(nanoflares) contribute in some way to coronal heating, they could be
too small and frequent to be resolved as independent events. In this
case, we would need to search for indirect evidence. 

The idea of this work is that, if the solar corona emission is
sustained by repeated nanoflares, locally the X-ray emission may not
be entirely constant but may show variations around the mean
intensity. So nanoflares may leave their signature on the light
curves. Many authors
\citep{1997ApJ...486.1045S,2000ApJ...541.1096V,2001ApJ...557..343K,2003PASJ...55.1025K,2008ApJ...689.1421S}
pointed out that a detailed analysis of intensity fluctuations of the
coronal X-ray emission could give us information on these smallest
flares. Following this hint we use this approach for the first time on
{\em Hinode} data, searching, with statistical analysis, for small but
systematic variability in noisy background light curves and their link
to coronal heating models.

\section{Instruments and Data}
The Japanese {\em Hinode} satellite was launched in 2006 
\citep{2007SoPh..243....3K}. One of the instruments it carries is the 
X-Ray Telescope (XRT), a high resolution grazing incidence telescope
\citep{2007SoPh..243...63G,2008SoPh..249..263K} that images the Sun 
through nine X-ray filters, using two filter wheels.

In our work we analyze the active region (AR) 10923, observed when it
was located close to the disk center on November 14, 2006. The data
set consists of $303$ images of $256 \times 256$ pixels, taken in the
Al\_poly filter with an average cadence of about 6~s, for a total
coverage of the observation of about $26$ minutes.

\begin{figure*}[t]       
\centering
\hspace*{-.5em}
	\includegraphics[width=6.5cm]{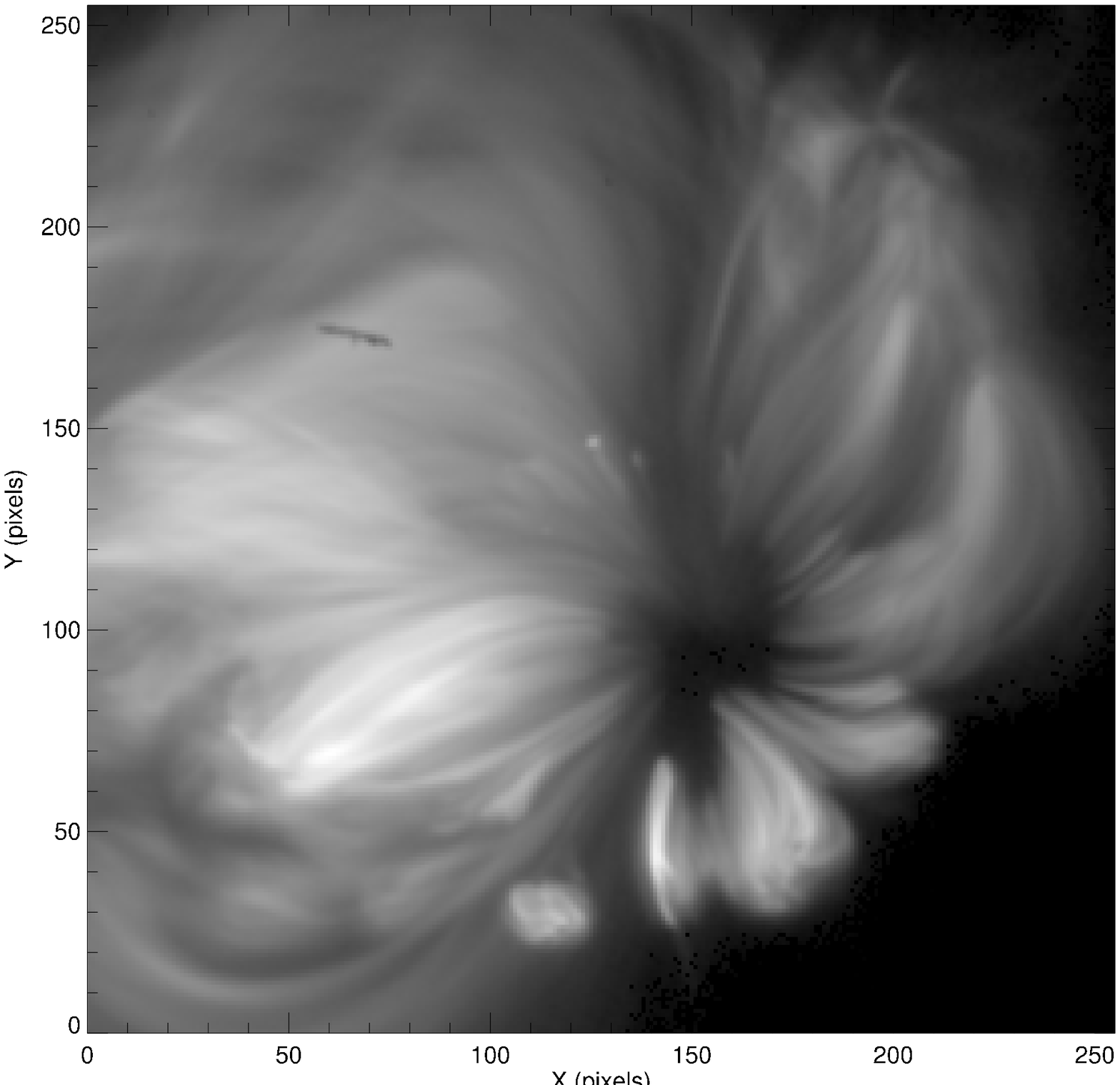}
	\includegraphics[width=6.3cm]{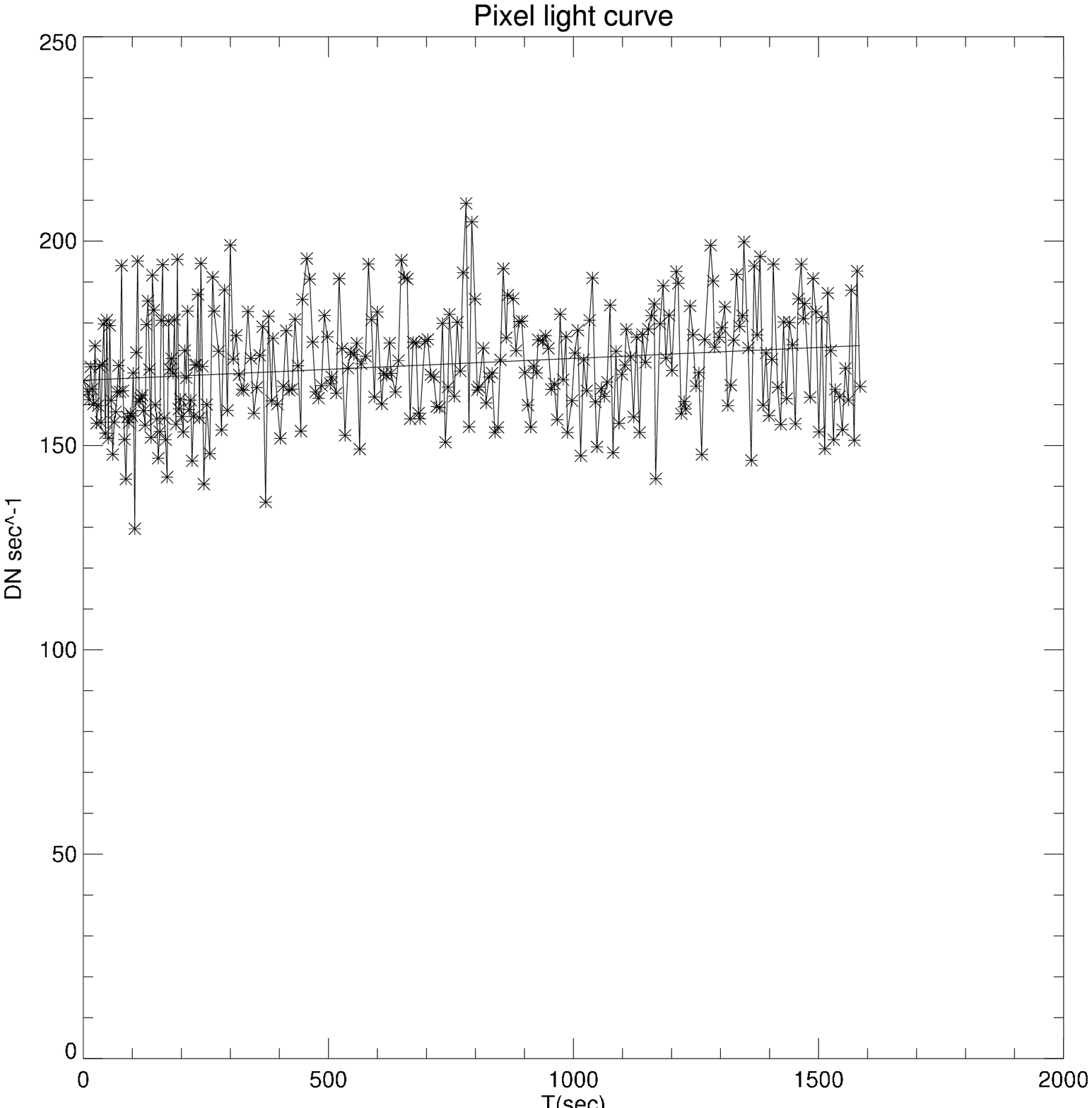}
\caption{{\em Left:} Average intensity map of the analyzed field of view 
($256 \times 256$ pixels), taken by {\em Hinode}/XRT on November 14,
2006, through the Al\_poly filter. {\em Right:} Example of pixel light
curve. The $y$-axis shows intensity in DN~s$^{-1}$, the $x$-axis is
time in s.}
\label{anda_3}
\end{figure*}

\begin{figure*}[t]       
	\centering
	\fbox{	
	\includegraphics[width=11.cm]{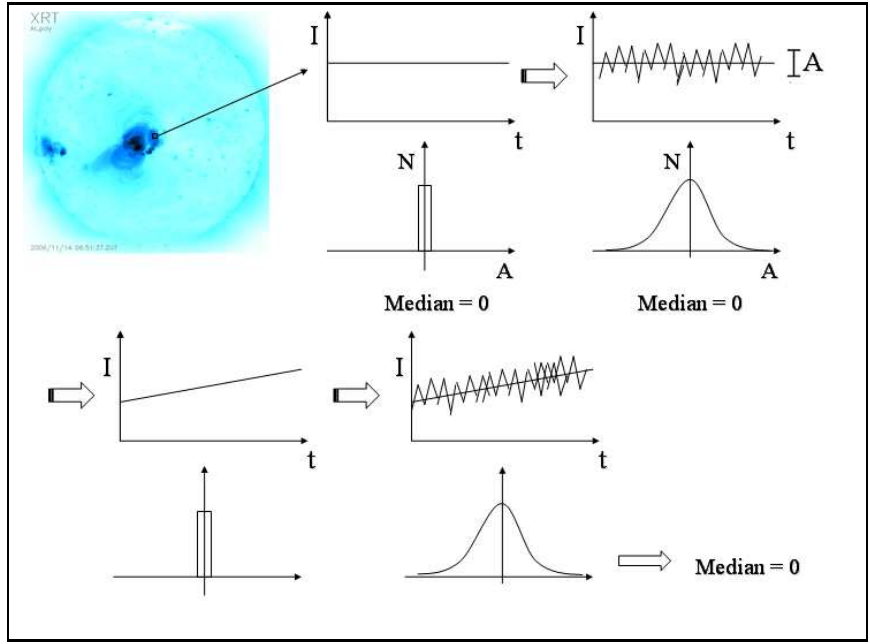}
	     }
\caption{Schematic concept of this work.  Consider the light curve 
of a valid pixel. If the count rate is constant, the histogram of 
fluctuations is trivial. If we add the expected photon noise and 
the count rate is high we expect a broad Gaussian distribution 
of the fluctuations. We expect the same distribution if the 
count rate is linearly changing.}
	\label{flutt}
\end{figure*}

\section{Data Analysis}
In this work we analyze the intensity fluctuations of a high cadence
XRT observation, searching for signal on top of statistical
fluctuations, and its link to coronal heating
model. Figure~\ref{flutt} is a flow chart which describes the concept
of our analysis. Let us consider the light curve of a pixel in the AR
and let us build the histogram of the amplitude of the fluctuations
with respect to the average emission trend. If the count rate is
strictly constant, the histogram of fluctuations is trivial. However,
fluctuations are intrinsic to the finite photon counting. If we add
the expected photon noise and the count rate is high, we expect an
almost Gaussian distribution of the fluctuations, symmetric around
zero, i.e., we expect an equal amount of positive and negative
fluctuations. We expect the same distribution if the count rate is
linearly and moderately changing instead of being constant.  Even for
a large curve with a constant rate we expect a broad distribution of
fluctuations due to photon noise. If the count rate is high the
distribution is approximately Gaussian. We expect exactly the same
distributions also if the count rate is linearly changing instead of
being constant, as sketched in Figure~\ref{flutt}.

Along this line, for each pixel we create the distribution of
intensity fluctuations of the light curve. We normalize the amplitude
of the fluctuations to the photon noise $\sigma_{\rm p}$, expected from
determining the original photons impacting on the detector from the
measured readout. The conversion from DN to photon counts is through
the conversion factor $K^{(2)}_i$ for the $i$-th filter-band:

\begin{equation}
 \sigma_{\rm p} = \sqrt{K^{(2)}_i(T)I_0},
\end{equation}
where $T$ is the electron temperature and $I_0$ is the measured DN.

The conversion factor is obtained as:
\begin{equation}
 K^{(2)}_i = \frac{\int{[(hc/\lambda)/(57\times3.65 {\rm eV})] \, P(\lambda,T) \, \eta_i(\lambda) \, {\rm d}\lambda}}{\int{P(\lambda,T) \, \eta_i(\lambda) \, {\rm d }\lambda}},
\end{equation}
where $P(\lambda,T)$ is the emissivity as a function of wavelength
$\lambda$ and electron temperature $T$, and $\eta_i(\lambda)$ is the
telescope effective area. For the moment, we assume a temperature
$\log T= 6.3$ \citep{1995ApJ...454..934K,2011SoPh..269..169N}.

\begin{figure}[!t]       
	\centering	
	\includegraphics[width=8cm]{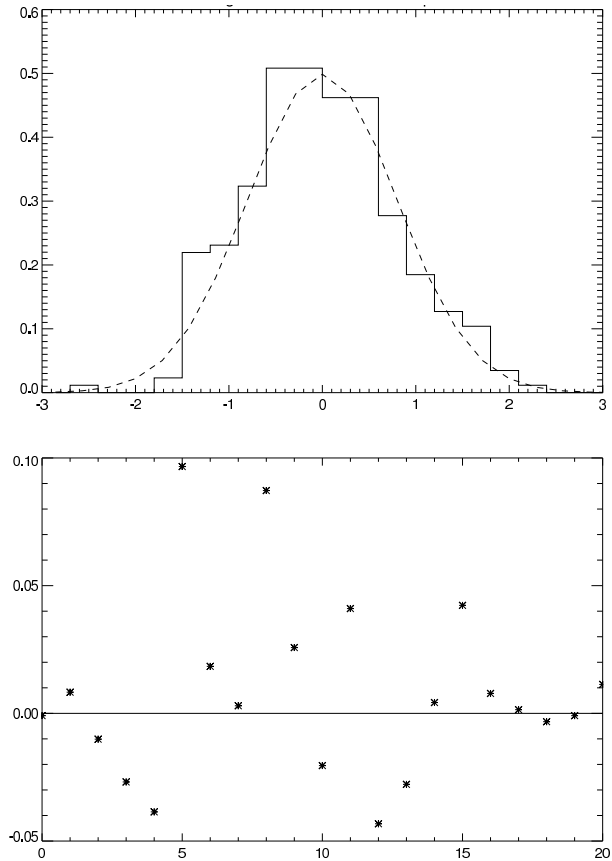}
\caption{{\em Top:} Distribution of the fluctuations of a pixel 
light curve, with respect to the linear fit to the light curve (solid
line) in comparison with a symmetric Gaussian-like distribution
(dashed line).  {\em Bottom:} Residuals of the fluctuations with
respect to the Gaussian curve.}
\label{histo_pix}
\end{figure}

Since we are interested in low amplitude systematic variations, we
first remove pixels with low signals and those showing all kinds of high
or slow amplitude systematic variations, such as:
\begin{itemize}
 \item Pixels with spikes due to cosmic rays
 \item Pixels that show micro-flares or other transient brightenings
 \item Pixels that show slow variations due, for instance, to local loop 
drift or loop motion.
\end{itemize}

At the end of this data screening we are left with about $56\%$ of
total pixels. The light curves of all these pixels can be fit with
linear regression. An an example of our analysis, in Fig.~\ref{histo_pix}
we show the result for one pixel (1,1).

\begin{figure*}[t]       
	\centering
	\fbox{
	\includegraphics[width=11.cm]{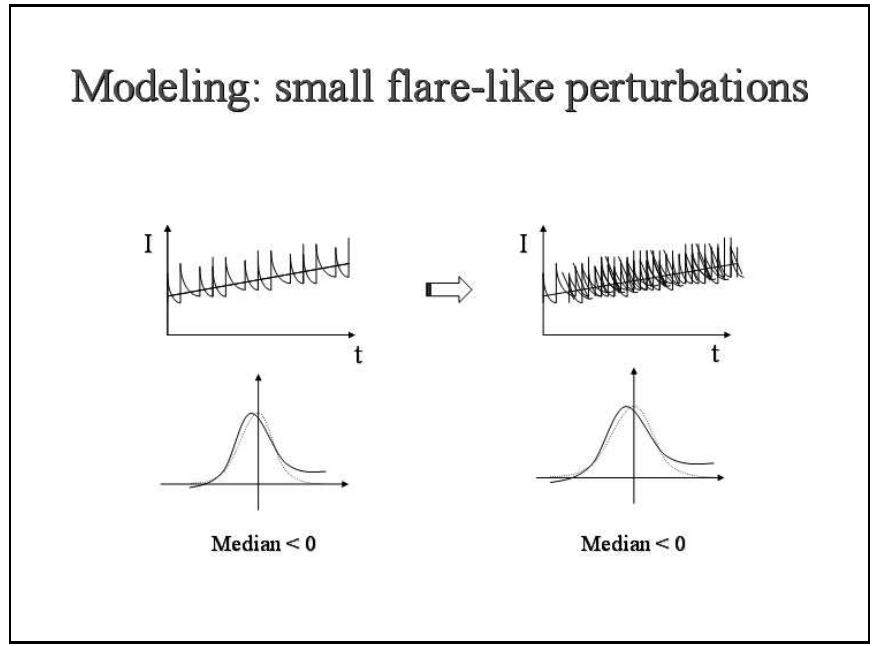}
	     }
\caption{Schematic concept of expectations from light curves 
perturbed by trains of small flare-like events. Although the mean
emission would not change, the distribution of the fluctuations would
become asymmetric, with a negative median even in the presence of
significant photon noise.}  \label{nanof}
\end{figure*}

\subsection{Results}
For each pixel we build the histogram of the X-ray intensity
fluctuations with respect to the linear fit
line. Figure~\ref{histo_pix} shows a sample distribution for one pixel
compared to a symmetric Gaussian distribution. By inspecting the
figure, we realize that there is a slight excess of the number of
negative fluctuations over the number of positive ones. This excess
makes the histogram slightly asymmetric toward the left side of the
distribution. The excess is present but not very significant on most
of the single pixel signal over the noise (as we can see in
Figure~\ref{histo_pix}).

From preliminary analysis, the excess becomes more significant when we
apply the analysis to selected subregions localized in different part
of the active region and even more when we sum over the whole region.

\section{Conclusions}
In summary, histograms of light curve fluctuations measured at high
cadence show asymmetric distributions with median lower than
zero. Asymmetries seem to appear at all scales from single pixels,
sub-regions, to the whole active region at high confidence level. We
propose that the asymmetry of the distributions of the fluctuations
could be compatible with random sequences of small fast rise and slow
decay pulses. This is typical of flare-like evolution
\citep{2010LRSP....7....5R}. Figure~\ref{nanof} sketches the concept
of this scenario: the presence of trains of small flare-like events
would make the distributions of fluctuations asymmetric in the
observed direction. The median would be negative because the light
curve will spend more time below than above the average trend,
i.e., the mean for a constant light curve.  Preliminary modeling using
Monte Carlo simulations appears to support this scenario. In our
opinion, if confirmed, these results would support a widespread
storm-of-nanoflares scenario. We emphasize that this analysis requires
high-cadence XRT observations. Improved diagnostics would be obtained
if at least a few images were available in different filter bands.

\acknowledgements
\begin{sloppypar}
{\em Hinode} is a Japanese mission developed and launched by 
ISAS/JAXA, with NAOJ as domestic partner and NASA and STFC (UK) as
international partners. It is operated by these agencies in
cooperation with ESA and NSC (Norway). ST, FR, and MM acknowledge
support from Italian Ministero dell'Universit\`a e Ricerca and Agenzia
Spaziale Italiana (ASI), contract I/023/09/0.
\end{sloppypar}

\bibliography{hinode4}

\end{document}